\documentclass[a4paper,fleqn]{cas-sc}
\usepackage{graphicx}
\usepackage{float}
\usepackage{subfig}
\usepackage{multirow}
\usepackage[numbers]{natbib}
\usepackage{pdfpages}
\def\tsc#1{\csdef{#1}{\textsc{\lowercase{#1}}\xspace}}
\tsc{WGM}
\tsc{QE}
\tsc{EP}
\tsc{PMS}
\tsc{BEC}
\tsc{DE}
\usepackage{amsmath}
\usepackage{graphics}
\usepackage{epsfig}
\usepackage{mathrsfs}
\usepackage{amsmath,amssymb,amsthm}
\usepackage{tikz}
\usepackage{flushend}

\usepackage{lineno}

\usepackage{multirow}
\usepackage{color}
\usepackage{array}
\newtheorem{lemma}{Lemma}

\usepackage[linesnumbered,ruled,vlined]{algorithm2e}
\usepackage{algpseudocode}
\usepackage{amsmath}

\begin{document}
\let\WriteBookmarks\relax
\def\floatpagepagefraction{1}
\def\textpagefraction{.001}
\shorttitle{Connected induced subgraph enumeration}

\shortauthors{Chenglong Xiao et~al.}

\title [mode = title]{An algorithm with a delay of $O(k\Delta)$ for enumerating connected induced subgraphs of size $k$}


%

\author[1]{Chenglong Xiao}

\address[1]{Department of Computer Science, Shantou University, China}
\author[1]{Chengyong Mao}
\author[1]{Shanshan Wang}
\cormark[1]

\cortext[cor1]{Corresponding author:Shanshan Wang, email:sswang@stu.edu.cn}

\begin{abstract}
The problem of enumerating connected subgraphs of a given size in a graph has been extensively studied in recent years. In this paper, we propose an algorithm with a delay of $O(k\Delta)$ for enumerating all connected induced subgraphs of size $k$ in an undirected graph $G=(V, E)$, where $k$ and $\Delta$ are respectively the size of subgraphs and the maximum degree of $G$. The algorithm requires a preprocessing step of $O(|V| + |E|)$ time to compute a depth-first search traversal order. The proposed algorithm improves upon the current best delay bound $O(k^2\Delta)$ for the connected induced subgraph enumeration problem in the literature.
\end{abstract}




\maketitle

\section{Introduction}

Enumerating the connected induced subgraphs of a given size from graphs has attracted much attention in recent years due to its wide range of applications. The subgraph enumeration algorithms have been used as backbone algorithms in many applications. Such application areas include bioinformatics \cite{1,2,3}, 6G core networks \cite{4}, machine learning \cite{5}, social networks \cite{6}, and car sharing systems \cite{7}, to name a few.

Previous works prove that the upper bound on the number of connected induced subgraphs of size $k$ in graph $G$ is $n \frac{(e\Delta)^{k}}{(\Delta-1)k}$, where $\Delta$ and $n$ are respectively the maximum degree of graph $G$ and the number of vertices in $G$ \cite{8,9}. Obviously,  the number of all connected induced subgraphs of a given size $k$ is exponentially large in $k$. This also implies that enumerating all connected induced subgraphs of a given size from a graph is a computationally challenging issue. Hence, an efficient algorithm is quite necessary. In the literature, a delay which is the maximal time that the algorithms spend between two successive outputs is usually used as a key indicator to evaluate subgraph enumeration algorithms \cite{10,11,12}. In this paper, we also use the upper bound of the delay for worst-case running time analysis of subgraph enumeration algorithms. Prior to reviewing the related works, we present the formal definition of the subgraph enumeration problem. Note that the definition was already given in \cite{11}.

\textit{Problem $GEN(G;k)$ }: Given an undirected graph $G = (V,E)$, the problem is to enumerate all subsets $X\subset V$ of vertices such that $|X|=k$ and the subgraph $G[X]$ induced on $X$ is connected.

Various works have been devoted to the problem $GEN(G;k)$ in recent years. Most algorithms for enumerating connected induced subgraphs found in the literature operate in a bottom-up fashion, beginning with a single-vertex subgraph and adding neighboring vertices successively to create larger connected induced subgraphs. Wernicke introduced an algorithm for enumerating all size-$k$ connected induced subgraphs \cite{1}, which labels each vertex and generates larger subgraphs by absorbing vertices from the set  $V_{Extension}$ into smaller subgraphs. These vertices in the set $V_{Extension}$ must satisfy two conditions: their assigned label must be larger than that of the vertex from which the larger subgraphs originate, and they can only be adjacent to the newly added vertex but not to any vertex in the parent subgraph. More recent work \cite{10} proposed a variant of Wernicke's algorithm by introducing a new pruning rule, which was found to enumerate all size-$k$ connected induced subgraphs with a delay of $O(k^{2}\Delta)$. This delay bound is currently the best delay bound in the literature.

Elbassuoni proposed two algorithms \cite{11}, named $RwD$ and $RwP$ respectively, which are designed specifically for enumerating connected induced subgraphs of size $k$ using the supergraph method with reverse search \cite{13}. The algorithms start by creating a connected induced subgraph with $k$ vertices and subsequently attempt to traverse every neighboring node in the supergraph where each solution corresponds to exactly one node of it by substituting a vertex in the current subgraph with a neighbor vertex. The $RwD$ algorithm has a delay of $O(k\,\mathrm{min}(n-k,k\Delta)(k(\Delta+\log{n}))$ and requires exponential space  $O(n+m+k|S|)$ for storing, where $|S|$ is the number of nodes in super graph $S$, $n$ is the number of vertices and $m$ is the number of edges. The $RwP$ algorithm requires only linear space $O(n+m)$  but with a larger delay of $O((k\,\mathrm{min}(n-k,k\Delta))^{2}(k(\Delta+\log{n}))$.

Another algorithm based on reverse search was proposed in \cite{2}. According to the algorithm, the vertex $s$ in a set of vertices with the smallest label is termed an anchor vertex. The utmost vertex is the one with the longest shortest path to the anchor vertex, denoted as $u$. The algorithm's objective is to expand subgraphs by incorporating each valid neighbor vertex $v$ that either has a distance farther from $s$ relative to the distance between $s$ and $u$ or has a lexicographically greater value than $u$, in cases where $v$ and $u$ share the same distance to $s$.

A more recent work was proposed in \cite{12}. The authors introduced a new neighborhood definition that two nodes ($X$ and $Y$) in the supergraph $\mathcal{G}$ are considered neighbors only if they share exactly $k-1$ vertices and the induced subgraph of their intersection ($X \cap Y$) is also connected. Three algorithms were proposed based on the new neighborhood. Two of the three algorithms slightly improve upon the current best delay bound $O(k^2\Delta)$\cite{10} in the case $k>\frac{n\log{\Delta}-\log{n}-\Delta+\sqrt{n\log{n}\log{\Delta}}}{\log{\Delta}}$ and $k>\frac{n^2}{n+\Delta}$ respectively at the expense of exponential space. It should be noted that the algorithms of \cite{11,12} preprocess the graph via a depth-first search (DFS) traversal for vertex ordering, which takes $O(|V|+|E|)$ time, while the algorithms of \cite{1,10} do not require such DFS ordering.

\section{Algorithm for enumerating connected induced subgraphs of size $k$}
In this section, we first present some necessary notations and preliminaries. Let $G=(V,E)$ be an undirected graph consisting of a vertex set $V=\{v_1,v_2,...,v_n\}$ and an edge set $E\subseteq V\times V$. The size of a subgraph denotes the number of vertices in the subgraph. The subgraph of $G$ induced by a vertex set $U\subseteq V$ is denoted by $G[U]$. The neighbors set of a vertex $v$ is expressed as $N(v):=\{u|\{u,v\}\in E\}$. A connected set is denoted by $S$ ($S \subseteq V$). The neighbor set of a connected set $S$ is denoted by $N(S):=\cup_{v\in S}N(v)\setminus S$. The delay of an enumeration algorithm is defined as the maximum amount of time spent between successive solution outputs. To ensure a polynomial delay bound, the time until the first solution is output and the time after the last solution is output must also be bounded by a polynomial function of the input size. In the rest of the paper, the size-$k$ subgraph refers to the connected induced subgraph of size $k$. A permitted neighbor of $S$ is a vertex in $N(S)$ that can be used to expand $S$. A forbidden neighbor denotes a vertex in $N(S)$ that cannot be used to expand $S$. Consistent with prior works \cite{1,10,15}, we also associate an enumeration tree with the proposed algorithm. To avoid ambiguity, each vertex in the enumeration tree is referred to as a node.

\begin{figure}[t]
  \centering
  \includegraphics[width=1.0\hsize=0.8]{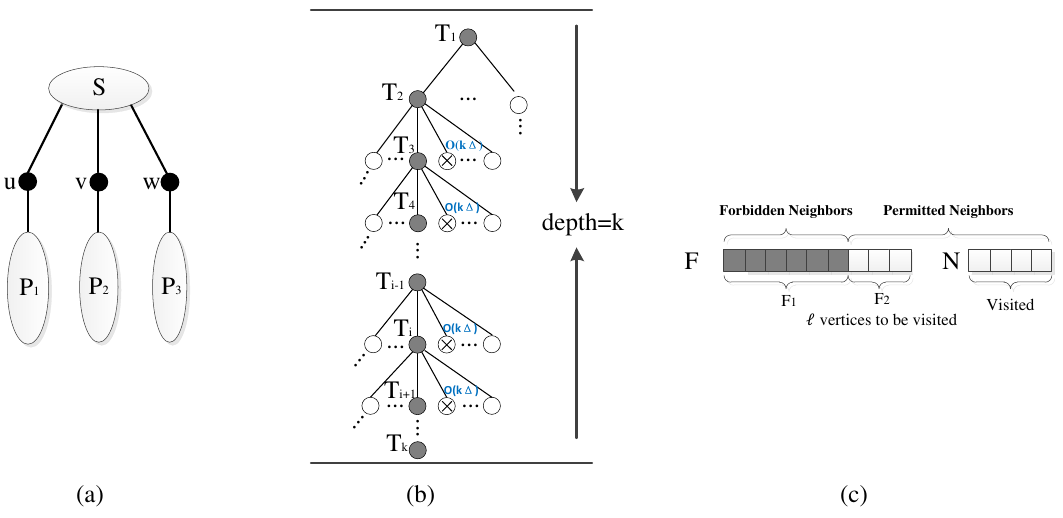}
  \caption{ (a) An example to illustrate the difference between the previous algorithms \cite{1,10,15} and the proposed algorithm.  $P_1$, $P_2$, and $P_3$ denote the set of vertices that are connected to $u$,$v$,$w$ by at least one path, respectively. For the sake of simplicity, we assume $P_1$, $P_2$, and $P_3$ do not share vertices. (b) Illustration of the case of $O(k^2\Delta)$ delay required by the previous algorithms. Each circle represents an enumeration node. The circle with a cross mark indicates that the corresponding enumeration node fails to find a solution. (c) An example of the set $F$ and the set $N$.}
  \label{f1}
\end{figure}

\subsection{Proposed algorithm}
Similar to the existing algorithms proposed in \cite{1,10,15}, the proposed algorithm incrementally expands a connected vertex set by incorporating neighboring vertices. The key differences between these existing approaches and our algorithm lie in the order in which vertices are visited and the strategy used to handle neighbor vertices.

\textit{Vertex visiting order.} Given a graph $G$, existing algorithms typically select an arbitrary vertex $v$, and recursively enumerate all size-$k$ subgraphs that include $v$. The vertex $v$ is then removed from $G$, and the process is repeated for another arbitrarily chosen vertex $u$ in $G \setminus \{v\}$. However, arbitrarily deleting a vertex from $G$ may create connected components of size smaller than $k$, necessitating a filtering step with time complexity $O(k^2\Delta)$ \cite{10}. In contrast, our algorithm first performs a DFS traversal of $G$, and then visits and removes vertices in the reverse order of this traversal. We show that removing vertices in reverse DFS order preserves the graph's connectivity, thereby avoiding the overhead of the additional filtering step.

\begin{lemma}
Let $G(V, E)$ be a connected undirected graph with at least two vertices, $u$ be the last vertex visited in a DFS traversal on $G$, then $G-\{u\}$ is connected.
\end{lemma}

\begin{proof}
Let \(T\) be the DFS tree generated by the traversal. Since \(G\) is connected and undirected, \(T\) is a spanning tree of \(G\). As \(u\) is the last vertex discovered in the DFS traversal, at the time of its discovery, all neighbors of \(u\) must have already been discovered. If any neighbor was undiscovered, the DFS would visit that neighbor next, contradicting that \(u\) is the last vertex discovered. Therefore, \(u\) has no children in \(T\), meaning \(u\) is a leaf in \(T\).

Removing a leaf from a tree results in a tree that remains connected. Thus, \(T - \{u\}\) is connected and spans all vertices in \(V \setminus \{u\}\). As \(T - \{u\}\) is a subgraph of \(G - \{u\}\) and \(T - \{u\}\) is connected, it follows that \(G - \{u\}\) is connected.

Therefore, for any connected undirected graph \(G\) with at least two vertices and any DFS traversal where \(u\) is the last vertex discovered, \(G - \{u\}\) is connected.
\end{proof}

By applying Lemma 1, we guarantee that the graph remains connected after each vertex is deleted in the reverse order of a DFS traversal. As a result, the costly $O(k^2\Delta)$ filtering step can be completely avoided.

\textit{The strategies for handling neighbor vertices.} Let $S$ be a connected vertex set with $|S|<k$, and let $N(S)$ denote its neighborhood. Both the existing algorithms and our proposed method iterate over each vertex in $N(S)$. In the previous approaches, once a neighbor is visited it is deactivated, that is, marked as forbidden for all subsequent recursive branches that consider the remaining unvisited neighbors. In contrast, our algorithm does the opposite: each visited neighbor is incrementally activated by marking it as permitted for all later branches.

To illustrate this distinction, consider the graph in Figure 1 (a). Here, the connected set $S$ has three neighbors $N(S)=\{u,v,w\}$, each linking to a set of vertices $P_1$, $P_2$, or $P_3$, respectively. Suppose all algorithms begin by visiting $u$. In existing methods, $v$ and $w$ are marked as permitted when enumerating all the size-$k$ subgraph containing $u$; by contrast, our algorithm forbids $v$ and $w$, pruning them from further consideration in this branch. Next, when the algorithms visit $v$, the previous approaches mark $u$ as forbidden, whereas our method retains $u$ as permitted. When the algorithms visit $w$, the previous approaches mark $u$ and $v$ as forbidden, while our method marks $u$ and $v$ as permitted.

Since the existing algorithms progressively exclude each visited neighbor from its sibling branches—thereby shrinking the pool of vertices available to extend the subgraph—they can immediately backtrack to the parent of the current enumeration node as soon as any recursive branch fails, without exploring the subsequent recursive calls.  To capitalize on this behavior, a new pruning rule was introduced in \cite{10}, which guarantees an $O(k^2\Delta)$ delay for these methods.

Figure 1(b) exemplifies the $O(k^2\Delta)$ delay incurred by existing algorithms. Consider an enumeration path $T_1, \ldots, T_k$ in the enumeration tree, where $T_k$ is a leaf node that yields a solution. Suppose the right sibling of any internal node $T_i$ (for $1 < i < k$) fails to find a solution. Each such failed sibling requires $O(k\Delta)$ time to process (comprising $k$ recursive calls, each taking $O(\Delta)$ time). Given the tree’s depth is $k$, backtracking from $T_k$ to the root $T_1$ accumulates a delay of $O(k^2\Delta)$, confirming the theoretical bound.

The deep backtracking observed in existing algorithms primarily arises from their strategy of progressively shrinking the set of candidate neighbors in sibling branches by excluding already visited vertices. Existing algorithms reduce the possibilities for subgraph expansion and often necessitate exploring multiple recursive branches that ultimately fail to find a solution in subsequent branches. In contrast, the proposed algorithm incrementally activates visited neighbors for use in subsequent sibling branches, that is, the subsequent sibling branches have more neighbors to expand. As a result, if the current branch yields a solution, the subsequent sibling branches are also guaranteed to find at least one solution, thereby avoiding unnecessary failed recursive paths. Moreover, our algorithm adds exactly one vertex to expand the current subgraph in each recursive call, ensuring that a solution can be found in at most $k$ recursive steps.

\begin{figure}[t]
  \centering
  \includegraphics[width=0.15\hsize=0.8]{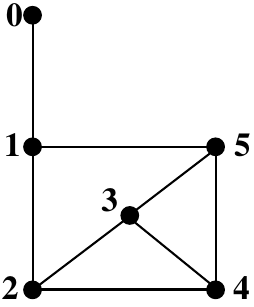}
  \caption{ A sample graph with 6 vertices. }
  \label{f2}
\end{figure}

Now, we describe the proposed algorithm. Similar to previous works, we assume that the input undirected graph $G(V, E)$ is connected in this work; otherwise, we can deal with each connected component individually. Algorithm 1 presents the pseudo-code of the proposed algorithm\footnote{The implementation of the proposed algorithm is available at https://github.com/miningsubgraphs/algorithms}. The algorithm receives an undirected graph $G = (V, E)$ and an integer $k$ as inputs and generates all connected induced subgraphs of size $k$ as outputs. It is composed of two procedures: $Main()$ and $Enumerate()$. The $Main()$ function first performs a DFS traversal on the input graph to obtain a vertex order (the DFS traversal can be performed in $O(|V|+|E|)$ time), and then iteratively initiates the enumeration from each vertex in reverse DFS order to maintain connectivity without additional filtering. The $Enumerate()$ procedure recursively explores and expands a connected subgraph by adding permitted neighbors one at a time, using a carefully designed neighbor handling strategy that incrementally activates neighbor vertices for future use. This approach ensures that each connected induced subgraph is output exactly once, with a delay of $O(k\Delta)$.

\textit{Data structures.} Before discussing the details of the proposed algorithm, we first define some important sets. A set $T$ is used to record all the visited vertices between two consecutive solutions. The vertices in $T$ include both the visited vertices neighboring the current set $S$ and the visited vertices that have a distance of more than one to $S$. With the set $T$, we can avoid redundant searching of visited vertices. For instance, suppose we first expand $S$ by adding vertex $u$ and $P_1$ (as shown in Figure 1(a)), and no solution is found during this step (i.e., $|S \cup \{u\} \cup P_1| < k$). We then proceed to the next iteration by expanding $S$ with vertex $v$ and $P_2$. 
Since $u$ and $P_1$ have already been explored, the vertices in $\{u\} \cup P_1$ are recorded in $T$ to prevent them from being re-examined. Once a new solution is found, the set $T$ is cleared.

Another important set is the neighbors of the set $S$.  We partition the neighbors of the set $S$ into two sets $F$ and $N$ respectively ($N(S)=N\cup F$ and $F\cap N=\emptyset$). The set $F$ consists of two parts: the front part stores forbidden neighbors, while the back part stores permitted neighbors that are yet to be visited. The set $N$ holds permitted neighbors that have already been visited in previous iterations or recursive calls. The purpose of maintaining $N$ is to avoid redundant exploration of previously encountered neighbors while preserving their information. Once a new solution is found, the vertices in $N$ are removed and appended to the end of $F$ as permitted neighbors that are yet to be visited. Figure 1(c) provides an illustration of the structure and roles of $F$ and $N$. In this paper, we use $F_1$ and $F_2$ to denote the front and back parts of $F$, respectively. In our algorithm, the sets $S$, $T$, $F$, and $N$ are implemented as global lists (or resizable arrays) and are initially empty lists. All operations such as appending or removing vertices are performed only at the end of each list.

In the algorithm, we use a global Boolean array $B$ with a fixed length of $|V|$ to record the existence of the vertices in $S$, $N$, and $F$ ($v\in N(u)\backslash (S \cup N \cup F)$), where each array index of $B$ corresponds to a vertex and the true value denotes that the vertex is in $S$, $N$ or $F$ (line 25, Algorithm 1). With the Boolean array $B$, we can perform the check if a vertex is in $S$, $N$, and $F$ in $O(1)$ time. Similarly, we also use a global Boolean array with a fixed length of $|V|$ to represent the existences of the vertices in $T$ . Thus, we can check if a vertex is in $T$ in $O(1)$ time (line 26). The initialization of the two aforementioned Boolean arrays takes $O(|V|)$ time.

\textit{Main loop.} In the main loop, vertices are processed in reverse DFS discovery order (i.e., from last discovered to first). For example, a DFS traversal starting at vertex 0 in Figure 2 yields the ordered set $V = \{ 0, 1, 2, 3, 4, 5\}$; the algorithm then processes vertices in the reverse order: 5,4,…,0. By Lemma 1, the remaining graph after removing each vertex $v$ in reverse DFS order remains connected, eliminating the $O(k^2\Delta)$ filtering in prior work \cite{1,10,15}.

We visit each vertex $v$ in $V$ from the back and add $v$ to form a one-vertex subgraph in lines 4,5 of the main loop. Next, the neighbors of $v$ that are not in $F$ are added to the end of a global list $F$ and marked as permitted neighbors ($|N(v)\backslash F|$ indicates the last $|N(v)\backslash F|$ vertices in $F$ are permitted neighbors and the remaining vertices of $F$ are forbidden neighbors, line 6). We call the recursive function $Enumerate(S, N, F,\ell)$ to enumerate all the size-$k$ subgraphs containing $v$ and excluding the forbidden neighbors in $F$. Then, we remove $v$ by appending it to $F$ as a forbidden vertex before entering the following iterations (line 8). If the number of unvisited vertices is less than $k$, the algorithm terminates (lines 10,11).

\begin{algorithm}[t]\footnotesize
   \caption{The Proposed Algorithm}
  \KwIn{An undirected graph $G=(V,E)$}
  \KwOut{A set of enumerated connected induced subgraphs of size $k$}
  \textbf{Procedure} $Main(G(V,E))$\\
  $F,N,S,T=\emptyset$\;
  V $\gets$ perform a DFS traversal on $G$\;
  \tcc{visit every vertex in $V$ from the back of $V$}
  \For{each vertex $v\in V$} 
  {
     $S=\{v\}$\;
     $F=F\cup (N(v)\backslash F)$; \textcolor{blue}{\Comment{append every vertex $u \in (N(v)\backslash F)$ to the end of $F$ }}\;
     $Enumerate(S,N,F,|N(v)\backslash F|)$\;
     append $v$ to the end of $F$ \textcolor{blue}{\Comment{implicitly remove $v$ from $G$ by marking $v$ as forbidden vertex}}\;  
     restore $S$ and $N$\;
    \If{$|V|-|F|<k$}
    {
       break\;
    }
  }
  
  \textbf{Procedure} $Enumerate(S,N,F, \ell)$\\
   hasSolution = False\;
   \If{$|S|=k$}
   {
      output $S$\;
      return True\;
   }
   \ElseIf{$|S|+|T|=k $}
   {
      output $S\cup T$ and clear $T$\;  
      hasSolution = True\;
   }
  \tcc{visit vertex from the back of $F$}
  \For{each vertex $u$ of the last $\ell$ vertices in $F$ } 
  {  
    remove last vertex $u$ from $F$\;
    append $u$ to the end of $S$\;
    $\ell'= 0$;  \textcolor{blue}{\Comment{the number of new permitted neighbors}}\;
     \If{ $|S|<k$ }
     {
       \For{each vertex $v\in N(u)\backslash (S \cup N \cup F)$}
       {
        \If{ $v\in T$ }
        {
           append $v$ to the end of $N$\;
        }
        \Else
        {
            append $v$ to the end of $F$\;
            $\ell'$++\;
        }
      }    
      \If{$hasSolution=True$}
      {
         $\ell' = \ell'+|N|$\;
         \For{each last vertex $w\in N$}
         {
            remove $w$ from $N$ and append $w$ to the end of $F$\;
         }
      }
    }
    \If{ $Enumerate(S,N,F,\ell')$ = True}
    {
          hasSolution = True\;
    }
    \Else
    {
          append $u$ to the end of $T$\;
    }
    restore $S$ and $N$\;
    append $u$ to the end of $N$\;
  }
  return hasSolution\;
\end{algorithm}

$Enumerate(S, N, F, \ell)$. The recursive function $Enumerate(S, N, F, \ell)$\footnote{Since the sets $S$, $N$ and $F$ are declared as global lists, it is not necessary to pass them as parameters in the function. Here, we present $Enumerate(\ell)$ with these sets to highlight the current state.} outputs all $k$-subgraphs containing $S$ and excluding the forbidden neighbors in the front part of $F$. The integer $\ell$ denotes the number of permitted neighbors at the end of $F$ that are yet to be visited.

In $Enumerate(S, N, F, \ell)$, we visit the last $\ell$ vertices in $F$ one by one from the back, remove each vertex from the end of $F$ and append it to the end of $S$ (lines 20-22). A global list $T$ is used to temporarily record all the visited vertices (not only the neighbors) during previous iterations and their recursions between two consecutive solutions. If the recursion called from the current enumeration node does not lead to finding a $k$-subgraph, we add the visited vertex to $T$ before returning to the parent enumeration node (line 38). 

If a $k$-subgraph is found, we clear $T$ ($|S|+|T|=k$, line 18). In other words, the algorithm continues to visit new vertices until the number of all the visited vertices involved in $S$ and $T$ is $k$. If the size of $S$ is less than $k$, for vertices in $N(u)$, we only visit the neighbor vertices that are not in $S$,$N$, and $F$ (line 25). The neighbor vertex $v$ may have been visited in previous iterations or their recursive calls and added to the set $T$ (line 26). To avoid losing any neighbors, we append $v$ to the end of $N$ in line 27, where $N$ tracks all neighbor vertices that have already been visited. If $v$ was not previously visited, we added it to the end of $F$ as a permitted neighbor yet to be visited (line 29).

Line 13 initializes a Boolean tag $hasSolution$ with False at the beginning of each recursive call. When at least one recursive call finds a $k$-subgraph, the Boolean variable $hasSolution$ is set to True in line 36. If a solution is found, we remove every last vertex from the end of $N$ and append it to the end of $F$ as a permitted vertex for use in subsequent branches (lines 33-34). The value of $\ell'$ is accordingly updated (line 32).  We may find a solution in two cases. The first case is when the number of vertices added to $S$ along the enumeration tree path reaches $k$ ($|S| =k$, line 14). The second case is when the combined number of vertices added  to $S$ along the enumeration tree path and the vertices added to $T$ during previous iterations and their recursions reaches $k$ ($|S|+|T|=k$, line 17).

\textit{Restoring the sets.} Since the sets \( S \), \( F \), and \( N \) are implemented as global lists, they may be modified by appending or removing vertices within the main loop and recursive function. Consequently, it is necessary to restore the contents of these sets before returning to the parent enumeration node or proceeding to the next iteration.

For the set $S$, since only one vertex is appended to the end in line 5 and line 22, we just simply remove the last vertex of $S$ to restore in line 9 and line 39. 

For the set $F$, we append vertices to the end of it in line 6 and lines 29, and 34. As these appended vertices will be removed by the child recursive call (line 21), the restoring of $F$ is not required. 

For the set $N$, we append the vertices to the end of it in lines 27 and 29 (the permitted neighbors added to $F$ in line 29 will be appended to $N$ in line 40 of the child recursive call.). In addition, we may remove every vertex in $N$ in case of $hasSolution=True$ (line 34). For the vertices appended in lines 27 and 29, we use a local counter to track the number of appended vertices and remove these vertices from the back of $N$. As for the vertices removed in line 34, since they are appended to $F$ and will be added back to $N$ in line 40 of the child recursive call, restoring for these vertices is not required.

\subsection{Case study}
Now, we illustrate how the proposed algorithm works by presenting an example of the enumeration tree of the algorithm on the sample graph presented in Figure 2, where the size of connected induced subgraphs is set to 5 (Figure 3). In the following case study, we assume that the algorithm first performs a depth-first search (DFS) traversal on the input graphs starting at vertex 0, producing an order $V = {0, 1, 2, 3, 4, 5}$. The algorithm visits vertices in the reverse of this order, i.e., $5 \rightarrow 4 \rightarrow 3 \rightarrow ... \rightarrow 0$.

Thus, we visit the vertex 5 first. The neighbors of the vertex 5 are appended to the end of $F$. These neighbors are marked as the last 3 vertices ($\ell=3$) to be visited in the recursive call of $Enumerate(\{5\},\{\},\{1,3,4\},3)$. Then, the call of $Enumerate(\{5\},\{\},\{1,3,4\},3)$ and its recursive calls are in charge of enumerating all the size-$k$ subgraphs containing the vertex 5. After these enumerations, the vertex $5$ is added to $F$ as a forbidden neighbor that cannot be included in the following iterations and their recursive calls. Note that we move the last three vertices from $F$ to $N$ inside the call of $Enumerate(\{5\},\{\},\{1,3,4\},3)$. The restoration of $F$ is not needed. We can directly add the vertex $5$ to $F$ as forbidden vertex. Next, the algorithm visits the vertex 4 in $Main()$, and the neighbor vertices of it are added to the end of $F$. The call of $Enumerate(\{4\},\{\},\{5,2,3\},2)$ and its recursive calls are in charge of enumerating all the size-$k$ subgraphs containing the vertex 4 and excluding the vertex 5. Until now, the vertex 5 and the vertex 4 cannot be included anymore. This indicates that the size of the remaining vertices in $G$ is less than $k=5$, thus the algorithm terminates.

\begin{figure}[t]
  \centering
  \includegraphics[width=0.96\hsize=0.8]{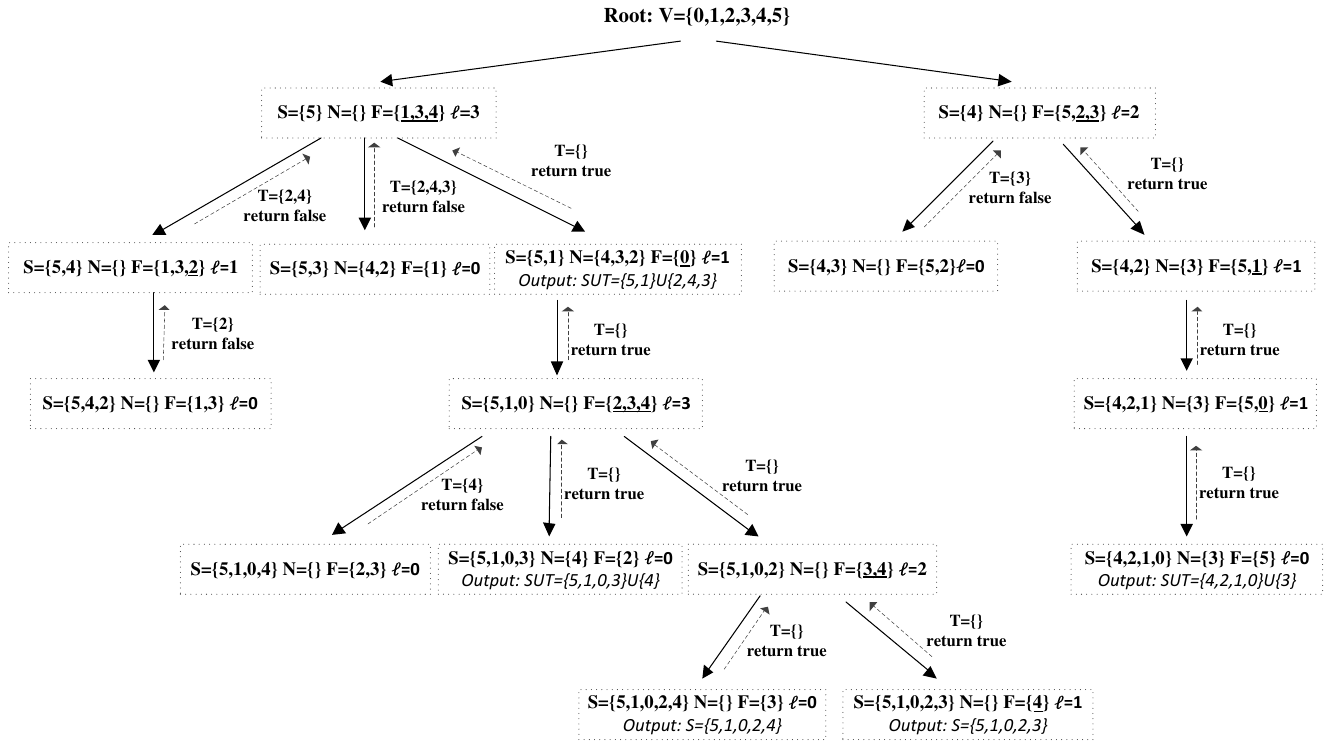}
  \caption{Enumeration tree of the proposed algorithm for enumerating all connected induced subgraphs of size 5 from the sample graph presented in Figure 2. Dashed rectangles of the tree indicate enumeration nodes. The enumerated connected induced subgraphs of size 5 are highlighted in italics. The vertices in $F$ that are underlined indicate that these vertices will be visited in the child node.}
  \label{f3}
  \vspace{-5mm}
\end{figure}

In the process of enumerating all the size-$k$ subgraphs containing the vertex 5, the algorithm visits the neighbor vertices of the current subgraph $S$ in a depth-first search way. For instance, starting from the enumeration node $[\{5\},\{\},\{1,3,4\},3]$, we remove the vertex 4 that is at the end of $F$ and append it to $S$ first. Then, the neighbor vertex 2 of the vertex 4 is visited and appended to $S$. When we arrive at the enumeration node $[\{5,4,2\},\{\},\{1,3\},0]$, the algorithm returns to the parent enumeration node as there are no more vertices that are marked as permitted neighbors in $F$. Since no solution is found, the algorithm appends each visited vertex to the end of $T$ by line 38 before returning the parent enumeration node  (e.g., before backtracking to $[\{5,4\},\{\},\{1,3,2\},1]$, vertex 2 is added to $T$; before backtracking to $[\{5\},\{\},\{1,3,4\},3]$, we add vertex 4 to $T$). In the enumeration node $[\{5\},\{\},\{1,3,4\},3]$, before visiting vertex 3, vertex 4 is appended to $N$ by line 40. 


When the algorithm returns to the enumeration node $[\{5\},\{\},\{1,3,4\},3]$ and visits the vertex 1 in $F$, we have $S=\{5,1\}$. As the neighbor vertex 2 of the vertex 1 was already explored and stored in $T$, we append it to the end of $N=\{4,3\}$ (which currently holds $\{4, 3\}$ after vertices 4 and 3 were added via line 40 in prior iterations). Thus, we have $N=\{4,3,2\}$. The neighbor vertex 0 is not in $T$, we append it to the end of $F$. In the following child recursive call, we have $|S\cup T|=5$ and output $S\cup T$ ($T$ is then cleared by removing all the elements in it). The local variable $hasSolution$ is set to true in this enumeration node. Then, we continue to visit the vertex $0$ in $F$. We remove it from the back of $F$ and append it to $S$. As $hasSolution$ is true, we remove every element from $N=\{4,3,2\}$ and append it to $F$ (lines 23-25). Thus, we have $N=\{\}$ and $F=\{2,3,4\}$ with $\ell=3$. The algorithm continues the aforementioned processes until the last solution $S=\{5,1,0,2,3\}$ is found. Then, the algorithm moves to find all the $k$-subgraphs containing the vertex 4 and excluding the vertex 5. It can be seen that each call of $Enumerate(S, N, F,\ell)$ in $Main()$ always terminates with a solution. 

\subsection{Correctness, delay and complexity of the algorithm}
The correctness, the delay, and the space of the proposed algorithm are provided and proven respectively as follows.

\begin{lemma}
Let $G(V, E)$ be a connected undirected graph, then the proposed algorithm correctly outputs all connected induced subgraphs of size $k$ from graph $G$ exactly once.
\end{lemma}

\begin{proof}
To prove the correctness of the proposed algorithm, it is sufficient to prove that the proposed algorithm outputs each connected induced subgraph of size $k$, and each connected induced subgraph of size $k$ is enumerated at most once.

\textbf{Firstly}, we show that if $G[C]$ is a connected induced subgraph of size $k$ with $C=\{u_1,u_2,...,u_k\}$, then $C$ is output by the proposed algorithm at least once. Let $\prec$ be the reverse DFS order used in the algorithm. Let $v_{\min} \in C$ be the last vertex in DFS traversal among all vertices in $C$ (i.e., the first in reverse DFS order). Thus, for all $v \in C \setminus \{v_{\min}\}$, we have $v_{\min} \prec v$. We prove by induction on the size $|C'|$ of the partial connected set $C' \subseteq C$ built during enumeration, that the algorithm will eventually construct $C$. $C'$ can be formed by $S\cup T$ or $S$. Here, $C'=S\cup T$ (where $T=\emptyset$ if $C'$ is built solely via $S$). Let $F_1$ and $F_2$ denote forbidden and the permitted vertices in $F$, respectively.

\textbf{Base Case ($|C'|=1$):} The main loop visits \( v_{\min} \), initializing \( C' = \{v_{\min}\} \) (line 5). All neighbors of \( v_{\min} \) in \( C \) are appended to \( F_2 \) (line 6). Since \( C \) is connected, \( |N(v_{\min}) \cap C| \geq 1 \), so there exists a neighbor \( v' \in N(v_{\min}) \cap C \) to expand \( C' \).

\textbf{Inductive Step ($|C'| = m < k$):} Assume \( C' \subset C \) is connected with \( |C'| = m < k\). Since \( C \) is connected, there exists \( u \in C \setminus C' \) such that \( u \in N(C') \). Consider three cases for \( u \):

\textit{Case 1: $u\in F_2$.}  If $u$ has not yet been visited, $u$ will be processed in the loop (line 20) and used to expand $C'$ by adding it to $S$ (line 22) or $T$ (line 38).

\textit{Case 2: $u\in N$.} If \( u \) was previously visited and stored in \( N \) by line 27, this implies $u \in T$. Since \( C' = S \cup T \), this contradicts \( u \in C \setminus C' \), as \( u \) would already be in \( C' \).  Thus, this case can be ruled out. If \( u \) was previously visited and stored in \( N \) by line 40, we consider two subcases:

1). A solution was found ($hasSolution = True$), by lines 31-34, all vertices in $N$ including $u$ are moved to $F_2$ as permitted vertices. Thus, $u$ can be used to expand $C'$.

2). No solution was found ($hasSolution = False$), by line 38, $u\in T$. This again contradicts with $u \in C \setminus C'$. Thus, this subcase can be ruled out.


\textit{Case 3: $u\in F_1$.} This can happen in two ways: 

1). It was visited before $v_{\min}$ and added as a forbidden vertex in the main loop by line 8. This can be ruled out because $v_{\min}$ is the first vertex in reverse DFS order. 

2). It was marked as a forbidden vertex during the construction of $C'$. According to the algorithm's neighbor handling rule, the newly marked forbidden vertices in an enumeration node remain only for its descendants. For subsequent sibling branches of that node, these forbidden vertices will become permitted vertices. Hence, $u$ is available to expand $C'$ in those sibling branches\footnote{Assume that $\{u, v\}$ are two permitted vertices currently under consideration, and that $v$ is visited first. In all descendant branches resulting from adding $v$, the vertex $u$ is marked as forbidden. However, in subsequent sibling branches, $u$ is a permitted vertex and can be used to expand the connected set.}.


In all cases, the algorithm eventually considers $u$ as a candidate for expanding $C'$, thus increasing $|C'|$ to $m+1$. By induction, the connected set $C$ will be constructed. Once $|S| = k$, or $|S| + |T| = k$, the algorithm outputs $C$ (lines 15 or 18).

\textbf{Secondly}, we prove that if $G[C]$ is a connected induced subgraph of size $k$, the proposed approach outputs $C$ at most once. Assume for contradiction that a subgraph \( C \) is output twice as \( C_1 \) and \( C_2 \). Since solutions are output when \( |S| = k \) (leaf node) or \( |S| + |T| = k \) (internal/leaf node), so \( C_1 \) and \( C_2 \) arise from either two distinct enumeration paths or the same enumeration path.  


\textit{Case 1.} Assume by contradiction that the same set $C$ is output twice via two distinct enumeration paths $P_1$ and $P_2$. These paths must diverge at some node in the enumeration tree. Let $T_i$ be the deepest node that is common to both paths and assume that $P_1$ precedes $P_2$ in the execution order. Two subcases arise for consideration: 

1). In the first case, $T_i$ is the root of the enumeration tree representing the function $Main( )$, suppose $P_1$ and $P_2$ choose $u$ and $w$ respectively (line 5). After choosing $u$, \(u\) is marked as a forbidden vertex for all subsequent branches (line 8), including \(P_2\). Consequently, no connected set output along \(P_2\) can contain \(u\). 

2). In the second case, $T_i$ is distinct from the root representing the function $Enumerate( )$, suppose $P_1$ and $P_2$ again choose vertices $u$ and $w$ respectively (line 22). After choosing $u$, all remaining vertices in $F_2$, including $w$, are marked as forbidden and cannot be used in any recursive subtree below. Therefore, any connected set output in the left subtree, including those along $P_1$, cannot contain $w$. 

Both subcases contradict the assumption that both $P_1$ and $P_2$ output the same set $C$.

\textit{Case 2.} We consider the case that $C_1$ and $C_2$ are enumerated by the same enumeration path. We further assume $C_1$ is enumerated before $C_2$, which implies that $C_1$ is output by an internal enumeration node in line 17 of the algorithm ($C_1=S\cup T$). If we track the reverse direction of the enumeration node outputting $C_2$, it has to eventually arrive at a single enumeration node $T_i$ that outputs $C_1$. In $T_i$, the enumeration path that outputs $C_2$ includes a vertex $v$ from the last $\ell$ vertices in $F$. However, according to the definition of $F$, every vertex in $F$ is a neighbor of $C_1$, that is, $F\cap C_1=\emptyset$. Thus, $v \notin C_1$. Therefore, $C_1\neq C_2$, we arrive at a contradiction.

Hence, each size-$k$ subgraph is enumerated exactly once.

\end{proof}

\begin{lemma}
   The delay of the proposed algorithm is $O(k\Delta)$, where $\Delta$ is the maximum degree of $G$.
\end{lemma}

\begin{proof}

\textbf{Firstly}, we show that the proposed algorithm requires at most $k$ calls of $Enumerate(S, N, F, \ell)$ to find a new solution. In $Main()$, we visit each vertex $v$ from the back of $V$ in terms of the reverse order of any depth-first search. Then, we call $Enumerate(S, N, F,\ell)$ to enumerate all the solutions containing $v$ and excluding the forbidden neighbors in $F$. The vertex $v$ is then marked as a forbidden neighbor that cannot be included in the subsequent calls of $Enumerate(S, N, F,\ell)$ in $Main()$. Since we visit the vertices in $V$ in reverse order of depth-first search, we can ensure that the remaining vertices ($V\backslash F$) are always connected. If the size of the remaining vertices is less than $k$, we terminate the enumeration. Otherwise, we know that there is at least one $k$-subgraph in the remaining vertices. Thus, each call of $Enumerate(S, N, F,\ell)$ in $Main( )$ leads to at least one solution. 

Let $T_i$ be an enumeration node in the enumeration tree of $Enumerate(S, N, F,\ell)$, we visit and remove every vertex $u_i$ of the last $\ell$ vertices in $F$ and enumerate all the solutions containing $u_i$ ($i=1,...,\ell$). In the enumeration of all the solutions containing $u_i$, the remaining vertices ($u_{i+1},...u_\ell$) in $F$ are marked as forbidden neighbors that cannot be included. For the subsequent recursive calls in $T_i$, $u_i$ is marked as a permitted neighbor. Since $u_i$ is a permitted neighbor for the subsequent recursive calls, all the vertices that can be added in the recursive call based on $u_i$ can also be added in these subsequent recursive calls. In other words, the subsequent recursive calls can include at least one more vertex (e.g., $u_{i+1}$) than the recursive call based on $u_i$. Hence, it is easy to know that if a recursive call in $T_i$ finds a solution, then all the subsequent recursive calls in $T_i$ can also find a solution.  Now, it is sufficient to show that we require at most $k$ calls of $Enumerate(S, N, F,\ell)$ for finding the first solution and spending between two consecutive solutions. Inside each call of $Enumerate(S, N, F,\ell)$, exactly one vertex is added to either $S$ or $T$. Thus, we can find the first solution or another solution with at most $k$ calls of $Enumerate(S, N, F,\ell)$.  

\textbf{Secondly}, we demonstrate the time required in each call of $Enumerate(S, N, F,\ell)$ without recursive calls. It is clear that each connected induced subgraph of size $k$ can be outputted in $O(k)$ time (lines 15, 18). As the length of $T$ is at most $k$, we can remove all the vertices from $T$ in $O(k)$ time (line 18). Note that outputting each connected induced subgraph of size $k$ and removing vertices from $T$ is only performed once between two consecutive solutions. The updates of $F$ and $S$ in each recursive call require $O(1)$ time (lines 21,22,29,34). Inside each recursive call, we check if each neighbor vertex of $u$ is not in $S$, $N$, and $F$ (line 25). To ensure $O(1)$ time of this check, we can use a global Boolean array $B$ with a fixed length of $|V|$ to record the existences of the vertices in $S$, $N$ or $F$, where each array index of $B$ corresponds to a vertex and its true value represents the vertex is in either $S$, $N$ or $F$. Similarly, we also use a global Boolean array with a fixed length of $|V|$ to record the existences of the vertices in $T$. 

We remove every vertex from the end of $N$ and append it to the end of $F$ when a solution is found in lines 33-34. It can be seen that these operations can be only called once between two consecutive solutions \footnote{Assuming a solution is currently found ($hasSolution=True$). In the next iteration, since $hasSolution=True$, the vertices in $N$ are removed and appended to $F$. Then, the $Enumerate$ function will be called. In all child recursive calls to $Enumerate$, the Boolean variable $hasSolution$ is initialized with False, and the operations of removing the vertices from $N$ and appending them to $F$ will not be performed. These operations are only executed when $hasSolution$ has been changed to True, meaning a new solution has been found. Thus, these operations can be only called once between two consecutive solutions.}. As $N$ stores a subset of vertices in $N(S)$ and $|N(S)|\leq k\Delta$, the time required for removing the vertices from $N$ to $F$ is at most $O(k\Delta)$. Since we have updated $S$, $N$, and $F$, we need to restore these three lists before returning to the parent enumeration node. To restore $S$, we only need $O(1)$ time to remove the last vertex from the end of it. For $F$, all vertices added to $F$ will be moved to $N$ by the child recursive call, the restoration of $F$ is not required. To restore $N$, we only need to remove the added vertices ($N(u)\backslash (S \cup N \cup F$) from the end of $N$. As $|N(u)|\leq \Delta$, the time for restoring $N$ in $Enumerate(S,N,F,\ell)$ is at most $O(\Delta)$. Therefore, the overall time required in each call of $Enumerate(S, N, F,\ell)$ is $O(\Delta)$ except for the time for removing every vertex from the end of $N$ and appending it to the end of $F$ (The operations for removing every vertex from the end of $N$ and appending it to the end of $F$ is only called once between two consecutive solutions).

Finally, we can conclude that the delay of the proposed algorithm is $O(k\Delta)$.

\end{proof}

\begin{lemma}
   The space complexity of the proposed algorithm is $O(|V|+|E|)$, where $|V|$ is the number of vertices in $G$, and $|E|$ is the number of edges in $G$.
\end{lemma}

\begin{proof}
It is clear that the space required for storing $G$ is $|V|+|E|$. According to the data structures used in the implementation, we know that $S$, $N$, and $F$ are declared as global lists. $S$ stores the set of vertices in the subgraph, and its maximum size is $k$. $N$ and $F$ store parts of the neighbor vertices of $S$ ($N \cup F=N(S)~and~N\cap F=\emptyset $). As $|N(S)|\leq min(k\Delta,|V|)$, then the maximum size of $N$ and $F$ is $|V|$. In the proposed algorithm, we also use a global list $T$ to record the vertices that are visited between two solutions. As we add at most one vertex to $T$ inside each call of $Enumerate(S, N, F,\ell)$ and we require at most $k$ calls to find a new solution, the maximum size of $T$ is $k$. In addition, a global Boolean array $B$ is used as an auxiliary array to check if a neighbor vertex is in $S$, $N$, and $F$. The length of $B$ is fixed to the size of the original graph $G$. The length of the auxiliary Boolean array for $T$ is also with a fixed length of $|V|$. Therefore, the space complexity of the proposed algorithm is $O(|V|+|E|)$.
\end{proof}

\section{Experimental Comparison}
In this section, we evaluate the proposed algorithm by comparing it to state-of-the-art algorithms. The algorithms under comparison are the $Simple$ algorithm \cite{1,10}, the $Simple-Forward$ algorithm \cite{10}, and the $VSimple$ algorithm \cite{15}. All these algorithms were implemented in the Java language. For the implementation of $Simple$ and $Simple-Forward$, we utilized the more efficient data structures proposed in \cite{15}. All experiments were conducted on an Intel Core i7 2.90GHz CPU with 8 gigabytes of main memory. The reported running times include the time taken to write the enumerated sets to the hard drive.

In our study, we utilized a set of graphs of varying sizes from the Network Repository. The dataset used in the experiments can be accessed at http://networkrepository.com \cite{16}. Consistent with prior studies, we divided the benchmark graphs into three categories: small graphs with fewer than 500 vertices ($n<500$), medium-sized subgraphs with 500 to 5,000 vertices ($500\leq n<5000$), and large graphs with 5,000 or more vertices ($n\geq 5000$). The characteristics of these benchmarks are detailed in Table 1, where columns $|V|$ and $|E|$ denote the number of vertices and edges for each graph, respectively. For the experiments, we established a running time threshold of 600 seconds. 

\begin{table*}[t]
\caption{Characteristics of Graphs Used in the Experiments}
\begin{center}
\begin{tabular}{llll}

\hline

 Size& Graph Name &     $|V|$ &    $|E|$   \\

\hline
Small&  ca-sandi\_auths &   86 &    124    \\

    & inf-USAir97 &   332 &    2126    \\

    &   ca-netscience & 379 &   914  \\

    &  bio-celegans &  453 &    2025 \\
\hline

Medium&      bio-diseasome &   516 &    1188 \\

     & soc-wiki-Vote &   889 &      2914 \\

     & bio-yeast &   1458 &    1948 \\

     & inf-power &   4941 &    6594    \\
\hline
  Large  &  bio-dmela &   7393 &    25569 \\

     & ca-HepPh &   11204 &    117619 \\

    &  ca-AstroPh &   17903 &    196972 \\

    &  soc-brightkite &   56739 &  212945 \\

\hline
\end{tabular}
\end{center}
\label{t2}
\end{table*}

Table 2 presents the cumulative running time for the proposed algorithm ($KDelta$), $Simple$, $Simple-Forward$, and $VSimple$ when enumerating all connected induced subgraphs with small cardinality ($k \in{2,3,4,5,6}$). Running time are reported in seconds. Given a 10-minute execution time threshold, some instances may exceed this limit (e.g., inf-USAir97 with $k=6$). The table includes only the total execution time for each algorithm that successfully generates the complete set of size-$k$ subgraphs within the threshold. All four algorithms produced identical sets of connected induced subgraphs for small cardinality $k$ as expected. The results indicate that $KDelta$ is the most efficient algorithm overall. Specifically, $KDelta$ outperforms the other three algorithms slightly for small and medium-sized benchmarks. For larger benchmarks, $KDelta$ achieves a 3.1X to 5.3X speedup over the fastest algorithm reported in the literature. Figure 4 illustrates the speedup of $KDelta$ over $VSimple$ across 12 benchmarks with small cardinality. During the experiments, it was observed that existing algorithms might not enumerate all size-$k$ subgraphs from large benchmarks (e.g., ca-HepPh, ca-AstroPh, and soc-brightkite with $k=4$) within the time threshold. However, $KDelta$ successfully generates all size-$k$ subgraphs from these large benchmarks within the specified threshold. Furthermore, the experimental results show that $VSimple$ marginally outperforms $Simple$ and $Simple-Forward$ in enumerating connected induced subgraphs of small cardinality $k$ in nearly all cases.

As the $TopDown$ algorithm proposed in \cite{15} is the most efficient algorithm for enumerating all connected induced subgraphs with large cardinality close to $|V|$ ($k \in\{|V|-3,|V|-2,|V|-1\}$), we only compare the proposed algorithm with $TopDown$. Table 3 shows the cumulative running time of the proposed algorithm ($KDelta$) and $TopDown$ for enumerating all connected induced subgraphs with large cardinality ($k \in\{|V|-3,|V|-2,|V|-1\}$). It can be seen that $TopDown$ outperforms $KDelta$ in all the instances. These results suggest that the top-down approach, which involves deleting vertices from the original graph, is more computationally efficient than the bottom-up method, which involves absorbing vertices into a single-vertex subgraph, when generating subgraphs of size $k$ that are close in size to the total number of vertices, $|V|$.

\begin{figure*}[h]
  \centering
  \includegraphics[width=0.54\hsize=0.8]{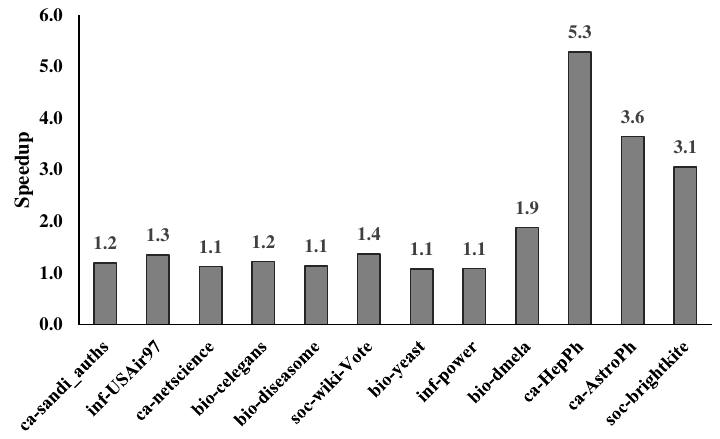}
  \caption{The speedup achieved by $KDelta$ over the state-of-the-art algorithm $VSimple$ on 12 benchmarks with small cardinality.}
  \label{f3}
   \vspace{-6mm}
\end{figure*}

\begin{table*}[t]
  \centering
  \caption{Running Time (in seconds) Comparisons of the Subgraph Enumeration Algorithms under Small Cardinality ($k \in{2,3,4,5,6}$). }

    \begin{tabular}{|l|c|c|c|c|c|}
    \hline
    \textbf{Benchmark} & Cardinality & \textbf{$KDelta$} & \textbf{$VSimple$} & \textbf{$Simple-Forward$}& \textbf{$Simple$} \\
    \hline
    ca-sandi\_auths & k=2,3,4,5,6  & \textbf{0.064}  & 0.077  & 0.079 & 0.082\\
    inf-USAir97 & k=2,3,4,5  & \textbf{22.672}  & 30.512  & 33.020  & 33.748\\
    ca-netscience & k=2,3,4,5,6  & \textbf{0.965}  & 1.087  & 1.067 & 1.126\\
    bio-celegans & k=2,3,4,5  & \textbf{60.292}  & 73.821  & 79.109  & 85.716\\
    bio-diseasome & k=2,3,4,5,6 & \textbf{3.695}  & 4.218  & 4.378 & 4.622\\
    soc-wiki-Vote & k=2,3,4,5,6  & \textbf{325.882}  & 446.154  & 511.771  &517.620\\
    bio-yeast & k=2,3,4,5,6  & \textbf{4.962}  & 5.341  & 5.550  &5.655\\
    inf-power & k=2,3,4,5,6  & \textbf{0.905}  & 0.987 & 1.039  & 1.052\\
    bio-dmela & k=2,3,4  & \textbf{6.451}  & 12.158  & 14.195  & 16.600\\
    ca-HepPh   & k=2,3  & \textbf{2.570}  & 13.591  & 14.023  &14.196\\
    ca-AstroPh & k=2,3  & \textbf{3.088}  & 11.269  & 14.251 & 14.860\\
    soc-brightkite & k=2,3  & \textbf{3.668}  & 11.231  & 11.474 & 17.263\\

    \hline
  
    \end{tabular}%
  \label{runtime}%

\end{table*}%

\section{Conclusion}
In this work, aiming at enumerating all connected induced subgraphs of a given size $k$, we presented a new algorithm with a delay of $O(k\Delta)$ and a space of $O(|V|+|E|)$. The delay bound of the proposed algorithm improves upon the current best delay bound of $O(k^2\Delta)$. To the best of our knowledge, the proposed algorithm is the first algorithm with a delay of $O(k\Delta)$ for the subgraph enumeration problem in the literature.

\begin{table*}[t]
  \centering
  \caption{Running Time (in seconds) Comparisons of the Subgraph Enumeration Algorithms under Large Cardinality Close to $|V|$ ($k \in\{|V|-3,|V|-2,|V|-1\}$). }

    \begin{tabular}{|l|c|c|c|}
    \hline
    \textbf{Benchmark} & Cardinality & \textbf{$KDelta$} & \textbf{$TopDown$} \\
    \hline
    ca-sandi\_auths & $k=|V|-3,|V|-2,|V|-1$  & 0.231  & \textbf{0.150} \\
    inf-USAir97 & $k=|V|-3,|V|-2,|V|-1$  & 88.419  & \textbf{40.216} \\
    ca-netscience & $k=|V|-3,|V|-2,|V|-1$ & 102.100  & \textbf{50.423}\\
    bio-celegans & $k=|V|-3,|V|-2,|V|-1$  & 345.043  & \textbf{182.811} \\
    bio-diseasome & $k=|V|-3,|V|-2,|V|-1$ & 295.620  & \textbf{151.907}\\
    soc-wiki-Vote & $k=|V|-2,|V|-1$  & 14.832  & \textbf{5.480}  \\
    bio-yeast & $k=|V|-2,|V|-1$  & 54.156  & \textbf{33.203} \\
    inf-power & $k=|V|-1$  & 1.590 & \textbf{0.602} \\
    bio-dmela & $k=|V|-1$  & 4.557 & \textbf{1.348}  \\
    ca-HepPh   & $k=|V|-1$   & 13.857  & \textbf{4.652} \\
    ca-AstroPh & $k=|V|-1$   & 44.635 & \textbf{23.192}  \\
    soc-brightkite & $k=|V|-1$   & 335.564  & \textbf{102.730}\\

    \hline
  
    \end{tabular}%
  \label{runtime}%

\end{table*}%

\section{Acknowledgement}
The authors would like to thank the anonymous reviewer for the careful reading and many valuable remarks, and the various grants from the National Natural Science Foundation of China (No.61404069), Scientific Research Project of Colleges and Universities in Guangdong Province (No.2021ZDZX1027), Guangdong Basic and Applied Basic Research Foundation (2022A1515110712 and 2023A1515010077), and STU Scientific Research Foundation for Talents (No.NTF20016 and No.NTF20017).

\end{document}